\documentclass[aps,prb,twocolumn,showpacs,preprintnumbers,superscriptaddress]{revtex4-2}

\usepackage{graphicx}
\usepackage{amsmath}
\usepackage{amssymb}
\usepackage{bm}
\usepackage{xcolor}
\usepackage[
  colorlinks=true,
  linkcolor=blue,
  citecolor=blue,
  urlcolor=blue
]{hyperref}

\usepackage{siunitx}

\newcommand{\EF}{$E_\mathrm{F}$}

\begin{document}

\title{Generation of an anomalous linearly dispersing spin-polarized band
       in Bi-based topological insulators}

\author{Matthias Kronseder}
\email{matthias.kronseder@ur.de}
\affiliation{Institute for Experimental and Applied Physics,
             University of Regensburg, 93053 Regensburg, Germany}

\author{Thomas Mayer}
\affiliation{Institute for Experimental and Applied Physics,
             University of Regensburg, 93053 Regensburg, Germany}

\author{Jan Min\'{a}r}
\affiliation{New Technologies -- Research Centre of the University of West Bohemia,
             Plze\v{n}, Czech Republic}

\author{Magdalena Marganska}
\affiliation{Institute for Experimental and Applied Physics,
             University of Regensburg, 93040 Regensburg, Germany}

\author{Hedwig Werner}
\affiliation{Institute for Experimental and Applied Physics,
             University of Regensburg, 93053 Regensburg, Germany}

\author{Florian Schmid}
\affiliation{Institute for Experimental and Applied Physics,
             University of Regensburg, 93053 Regensburg, Germany}

\author{Rebeca D\'{i}az-Pardo}
\affiliation{School of Natural Sciences, Department of Physics,
             Technical University of Munich, 85748 Garching, Germany}

\author{Ivana Vobornik}
\affiliation{Elettra-Sincrotrone Trieste, Trieste, Italy}

\author{Jun Fuji}
\affiliation{Elettra-Sincrotrone Trieste, Trieste, Italy}

\author{Cornelia Streeck}
\affiliation{Physikalisch-Technische Bundesanstalt, Berlin, Germany}

\author{Alexander Gottwald}
\affiliation{Physikalisch-Technische Bundesanstalt, Berlin, Germany}

\author{Hendrik Kaser}
\affiliation{Physikalisch-Technische Bundesanstalt, Berlin, Germany}

\author{Bernd K\"{a}stner}
\affiliation{Physikalisch-Technische Bundesanstalt, Berlin, Germany}

\author{Christian H. Back}
\affiliation{School of Natural Sciences, Department of Physics,
             Technical University of Munich, 85748 Garching, Germany}
\affiliation{Center for Quantum Engineering (ZQE),
             Technical University Munich, 85748 Garching, Germany}

\date{\today}

% ------------------------------------------------------------
\begin{abstract}
We report the generation of an anomalous linearly dispersing, spin-polarized band in Bi-based topological insulator (TI) thin films, induced by soft Ar-ion bombardment followed by annealing. This extra band---which we call the anomalous linearly dispersing state (ALS)---is superimposed on the regular band structure including the topological surface state (TSS), spans an unusually large energetic range of up to ${\sim}\,\SI{650}{\milli\electronvolt}$ at the $\Gamma$-point, and appears near the Fermi energy. Spin-resolved measurements indicate spin-momentum locking with a helicity \emph{opposite} to that of the regular TSS. The Fermi velocity of the ALS, $v_\mathrm{F} = (5.1\pm 0.4)\times 10^{5}\,\frac m s$, is indistinguishable from that of the regular TSS, $(5.3\pm 0.5)\times 10^{5}\,\frac m s$. The observation is reproducible across samples of varying thickness and was confirmed at two independent synchrotron radiation facilities. We discuss different mechanisms for the physical origin of the observed ALS including sputtering-induced TSS relocation, bi-layer formation by,e.g., chalcogen removal, and high-index surface relocation.
\end{abstract}


\maketitle

% ============================================================
\section{Introduction}
\label{sec:intro}

In Bi-based three-dimensional topological insulators (TIs), topological surface states (TSS) arise at interfaces with topologically trivial materials and exhibit spin-momentum locking, making them robust against perturbations that preserve time-reversal symmetry~\cite{Hasan2010,Qi2011,Zhang2009}. In combination with a bulk insulating gap, these properties make TIs attractive for spin-charge interconversion~\cite{Mellnik2014,Sanchez2016} and related spintronic devices. However, in practice the Bi-based TI family is prone to incorporate all kinds of defect states. Screening of excess charges originating from vacancy defect states induces pronounced electrostatic potential fluctuations. In thin films, these fluctuations lead to a reduced mobility of the TSS. In films thicker than $60$--$80\,\si{\nano\metre}$, three-dimensional charge puddles form as the bulk band gap is of only ${\sim}\,250$--$300\,\mathrm{meV}$~\cite{Bomerich2017}. These charge puddles lead to a large conductivity of the bulk fully masking the TSS and their exceptional properties with a conventional metallic behavior. Besides these defect state related problems, two other properties of the Bi-based TI-family are detrimental regarding applications. The first is the rather small density of states (DOS) of the TSS for all energies within the band gap which is even zero at the Dirac point. Hence, for many applications the Fermi energy needs to be away from the Dirac point but well within the band gap. Another disadvantageous side effect of the rather small energy gap is the influence of bulk bands on the dispersion relation of the TSS. Besides a, in general, non-zero curvature of the TSS bands away from the Dirac point, the warping effect belongs to this category imposing a hexagonal symmetry to the TSS-band dispersion in reciprocal space \cite{Fu2009}. 

We report the unexpected generation, via soft Ar-ion bombardment and annealing, of an anomalous linearly dispersing state (ALS) in $({\rm Bi}_{1-x}{\rm Sb}_x)_2({\rm Te}_{1-y}{\rm Se}_y)_3$ (BSTS)/Bi$_2$Se$_3$ heterostructures. The ALS exhibits a linear dispersion spanning up to ${\sim}\,\SI{650}{\milli\electronvolt}$---far exceeding the bulk band gap of the parent material---spin-momentum locking with helicity opposite to the regular TSS, and a Fermi velocity indistinguishable from that of the TSS. The near-identical Fermi velocities represent a strong constraint on proposed explanations. While the ALS may be topologically non-trivial in origin, its definitive classification awaits future theoretical and experimental investigation. We report the phenomenology and discuss possible origins.

% ============================================================
\section{Experimental}
\label{sec:experimental}

The investigated TI films are heterostructures consisting of two
quintuple layers (QLs) of Bi$_2$Se$_3$ (BS) followed by $d_\mathrm{BSTS}$
QLs of $({\rm Bi}_{1-x}{\rm Sb}_x)_2({\rm Te}_{1-y}{\rm Se}_y)_3$ (BSTS),
grown by molecular beam epitaxy on SrTiO$_3$(111) substrates.
Since BS is always n-type and BSTS was tuned to be slightly p-type with $(x|y) = (70|90)$, the n-p heterostructure design~\cite{Mayer2021,Eschbach2015} suppresses
the bulk carrier concentration and produces band bending along the
growth direction that enables determination of the Fermi level position
within the upper TSS~\cite{Danilov2021}. 
Samples were protected during ex-situ transfer to the ARPES facilities
by an (8\,nm)Te/(50\,nm)Se capping layer.

\begin{figure}[t!]
\centering
\includegraphics[width = 0.95\columnwidth]{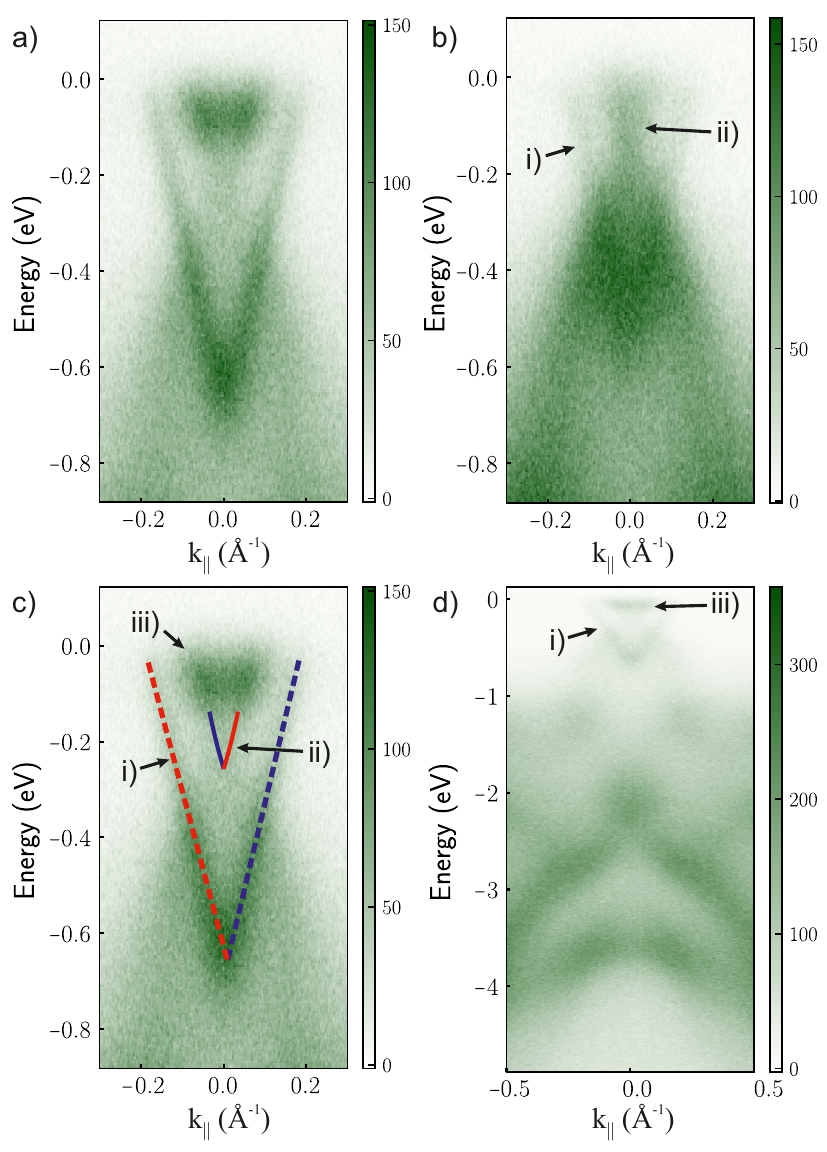}
\caption{a) ARPES scan with $E_\mathrm{ph}= 36$eV and \EF\ set to 0eV from a sample with (2)BS/(27)BSTS after a sputtering-heating cycle. The scan shows a faint TSS band structure in the background around -0.2eV. The main feature is, however, the linearly dispersing band starting at around -0.65eV (ALS), and the extra features close to \EF. b) The sister sample with nominally the same stoichiometry and thickness was treated similarly and shows also the linearly dispersing bands spanning this time only an energetic range of ~0.55eV, while the extra features close to \EF\ is not seen. c) same plot as in a) but highlighting the different features: i) ALS, ii) ordinary TSS, iii) extra bands close to \EF. d) large energy ARPES scan from sample shown in a) spanning most of the valence band structure and showing again both features i) and iii) around \EF.  }\label{fig:fig1}
\end{figure}

\begin{figure*}
\centering
\includegraphics[width = 0.9\textwidth]{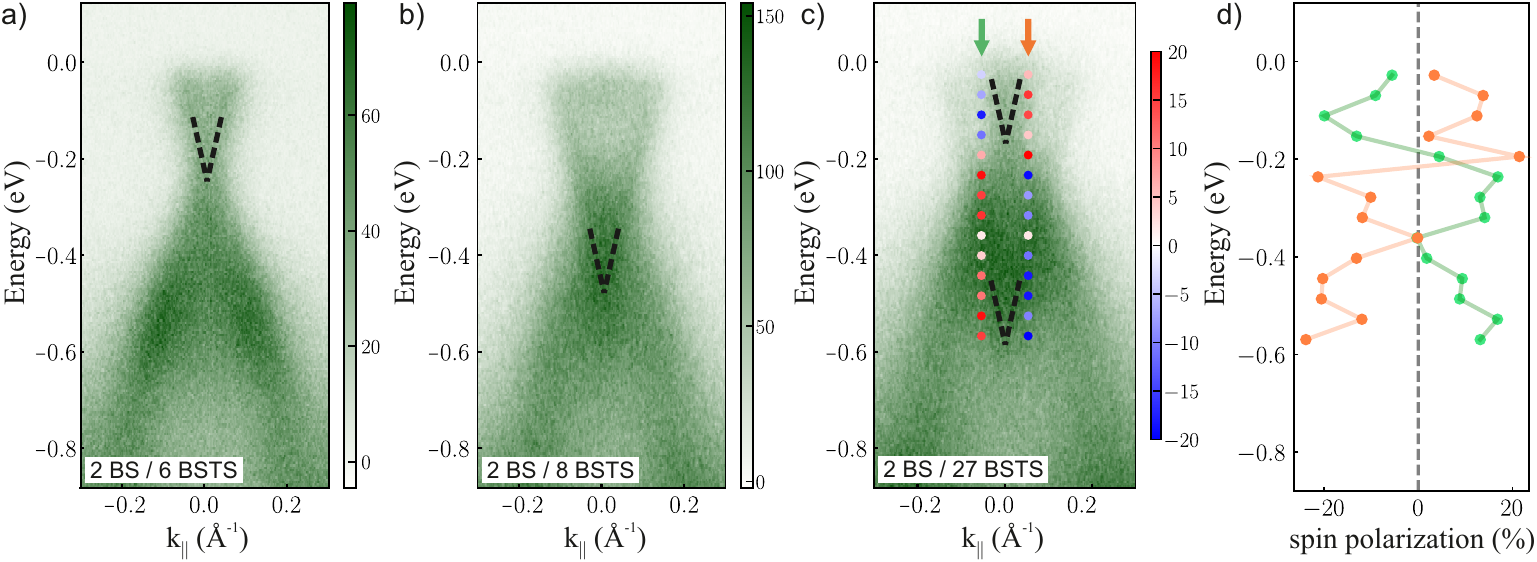}
\caption{a-c) ARPES scans of samples with (2)BS and different thicknesses of the BSTS layer after sputtering-heating cycles: a) for a $d_\mathrm{BSTS}=6$QL, b) $8$QL and c) $27$QL. For all plots the onset of the Dirac cone is highlighted by a dashed cone with the same size in all plots. In d) spin-polarization scans for two different in-plane wave vectors $k_\mathrm{\parallel}=\pm0.06\mathrm{\AA^{-1}}$ is plotted, $k_\mathrm{\parallel}<0$ in green, $k_\mathrm{\parallel}>0$ in orange for the sample with $d_\mathrm{BSTS}=27$QL. The same data is also overlaid in c) at the corresponding wavevectors and energies (marked with green and orange arrows) while the polarization in \% is given by the colorbar in c).  }\label{fig:fig2}
\end{figure*}

The capping layer was removed by annealing at $250$--$280\,{}^\circ\mathrm{C}$ for 30 minutes in ultra-high vacuum. Successful decapping was verified by ARPES: for $d_\mathrm{BSTS} > 5\,\mathrm{QL}$ the Fermi energy lies within the bulk band gap after decapping. Core-level spectra up to ${\sim}\,\SI{70}{\electronvolt}$ binding energy (covering Sb-4d, Bi-5d, Se-3d, Te-4d doublets) showed no stoichiometric changes upon further annealing at these temperatures. After this decapping procedure, a treatment cycle is applied to the samples, consisting of soft Ar-ion bombardment ($250$--$500\,\si{\volt}$ acceleration voltage, $\SI{10}{\micro\ampere}$, $5$--$10\,\si{\minute}$) followed by annealing at $270$--$280\,{}^\circ\mathrm{C}$ for up to 40\,minutes; this is hereafter referred to as the \emph{sputter-heating cycle}. To confirm that the observations are intrinsic to the sample and instrument-independent, the full procedure was independently replicated at Elettra Synchrotron (Trieste, Italy) and at the Metrology Light Source of the Physikalisch-Technische Bundesanstalt (Berlin, Germany); the ALS was observed at both facilities but all data shown in the main article are taken at the Elettra Synchrotron, data taken at the Metrology Light Source are shown in the Supplemental Material.

% ============================================================
\section{Results}
\label{sec:results}

\subsection{ARPES after the sputter-heating cycle}

Figure~\ref{fig:fig1}(a) shows an ARPES scan ($E_\mathrm{ph} = \SI{36}{\electronvolt}$)
of a (2)BS/(27)BSTS sample after the sputter-heating cycle.
Three distinct features are visible, highlighted in panel~(c):
\begin{enumerate}
  \item[(i)] A pair of linearly dispersing bands (red/blue dashed lines)
    spanning ${\sim}\,\SI{650}{\milli\electronvolt}$ with a crossing point
    near $-\SI{650}{\milli\electronvolt}$, which we call the \emph{anomalous
    linearly dispersing state} (ALS).
  \item[(ii)] The regular TSS (red/blue solid lines), with its Dirac point
    at ${\sim}{-}\SI{220}{\milli\electronvolt}$.
  \item[(iii)] A pronounced feature within a ${\sim}\,\SI{100}{\milli\electronvolt}$
    window around the Fermi energy.
\end{enumerate}

A nominally identical sample treated with an annealing temperature $10$--$30\,\si{\kelvin}$ lower [Fig.~\ref{fig:fig1}(b)] shows the same features (i) and (ii), but both are shifted to higher binding energy: the ALS crossing appears near $-\SI{500}{\milli\electronvolt}$ and the TSS Dirac point near $-\SI{150}{\milli\electronvolt}$. We suspect that feature iii) is absent because it is energetically related to feature i), and since feature i) occurs $-\SI{150}{\milli\electronvolt}$ higher in energy, feature iii) is unoccupied.

The simultaneous shift of features (i)--(iii) between nominally identical samples treated at slightly different temperatures strongly suggests an electronic coupling between the ALS and the bulk band structure including the regular TSS. Figure~\ref{fig:fig1}(d) shows a large-energy ARPES scan up to $\SI{5}{\electronvolt}$ below $E_\mathrm{F}$: the intact valence band structure of BSTS is visible together with the superimposed ALS and feature (iii), while the TSS is barely discernible, indicating that the ALS overlays the normal band structure.

\subsection{Thickness dependence}

The ALS crossing point seem to depend systematically on the BSTS layer thickness [Fig.~\ref{fig:fig2}(a--c)]: it shifts from ${\sim}{-}\SI{230}{\milli\electronvolt}$ for $d_\mathrm{BSTS} = 6\,\mathrm{QL}$ to ${\sim}{-}\SI{450}{\milli\electronvolt}$ for $8\,\mathrm{QL}$ and ${\sim}{-}550$ to $-\SI{650}{\milli\electronvolt}$ for $27\,\mathrm{QL}$, while the bulk valence band background remains essentially unchanged. This thickness dependence implies a sensitivity of the ALS to the heterostructure band alignment, distinct from simple bulk-band coupling. In the $27\,\mathrm{QL}$ sample two crossing points are discernible, suggesting the simultaneous presence of the ALS and a modified TSS. Core-level spectra measured before and after the sputter-heating cycle were inconclusive: the Sb:Bi and Te:Se ratios remained almost constant within the experimental uncertainty.

\subsection{Spin-polarization and Fermi-velocity}

Spin-polarization measurements were performed at two in-plane
wave vectors $k_\parallel = \pm 0.06\,\si{\per\angstrom}$
[Fig.~\ref{fig:fig2}(c,d)].
The two traces show opposite behavior and change sign near the Dirac
point of the regular TSS (feature ii), confirming spin-momentum locking
for the TSS.
Despite the lower spectral weight of the ALS, the data indicate that
the ALS also exhibits spin-momentum locking, but with a helicity
\emph{opposite} to that of the regular TSS.
The Fermi velocity of the regular TSS is
$v_\mathrm{F} = (5.3\pm 0.5)\times 10^{5}\,\frac m s$
and that of the ALS is
$v_\mathrm{F} = (5.1\pm 0.4)\times 10^{5}\,\frac m s$,
indistinguishable within experimental uncertainty.

% How have these values for the uncertainty of the Fermi velocity been determined (from Fig. 2 ?) ? I tried to fit different slopes and could determine that the given range any meaningful fit is covered. 

\subsection{Surface morphology}

AFM measurements [Fig.~\ref{fig:fig3}] reveal a significant morphological change after the sputter-heating cycle. Untreated surfaces show rounded quintuple-layer islands typical of quaternary BSTS growth. The scans shown in the upper row are provided for reference only. They depict the characteristic surface morphology typically observed immediately after MBE growth and are representative of the samples presented in the lower panels. After treatment, surfaces exhibit irregular island shapes with substantially higher roughness and a loss of long-range structural order. For thick samples ($27\,\mathrm{QL}$) the disordered region appears to extend over several QLs [Fig.~\ref{fig:fig3}(d)], while for thin samples the disorder is confined to the first $1$--$2\,\mathrm{QL}$ [Fig.~\ref{fig:fig3}(c)]. 

% ============================================================
\section{Discussion}
\label{sec:discussion}

The ALS has several salient properties that distinguish it from previously reported \cite{xia2009,hsieh2009} features in the Bi-based TI family: (1) a linear dispersion spanning up to ${\sim}\,\SI{650}{\milli\electronvolt}$, well exceeding the ${\sim}\,250$--$\SI{300\,}{\milli\electronvolt}$ bulk band gap; (2) a Dirac-like crossing at the $\Gamma$-point; (3) spin-momentum locking with helicity opposite to the regular TSS; and (4) a Fermi velocity indistinguishable from that of the regular TSS, despite the two bands coexisting. We now evaluate candidate explanations.

\subsection{Trivial Rashba state}

An important alternative is whether the ALS could be a trivial Rashba spin-split two-dimensional electron gas (2DEG) arising from inversion-symmetry breaking at the modified surface. Rashba 2DEG states are known to coexist with the TSS in Bi$_2$Se$_3$-family materials under band-bending conditions~\cite{Benia2011}, and spin-ARPES has confirmed that their helicity is opposite to the TSS~\cite{Jozwiak2016} consistent with our observation. However, Rashba 2DEG states exhibit a parabolic dispersion with a characteristic splitting at finite $k_\parallel$. The resolution in our data does not allow for a conclusive determination of a non-zero splitting at the $\Gamma$-point. However, the strict linearity of the ALS over $\SI{650}{\milli\electronvolt}$ makes parabolic Rashba-states very unlikely as reason for the ALS, but we cannot fully exclude this scenario, since the limited $k$-resolution of our spin measurement precludes a conclusive test.

\subsection{TSS relocation and preferential sputtering}

Queiroz \textit{et al.}~\cite{Queiroz2016} demonstrated for Bi$_2$Se$_3$ that noble-gas sputtering introduces strong dilute (unitary) disorder in the near-surface region, which pushes the TSS to deeper quintuple layers, burying it below a surface Anderson insulator. The ARPES signal then shows a weaker but still sharp Dirac cone located at the interface between the disordered surface region and the ordered bulk~\cite{Schubert2012}. This mechanism may account for our feature (ii)---the partially suppressed regular TSS. However, sputtering-induced relocation as in Ref.~\cite{Queiroz2016} produces one TSS, relocated, rather than two coexisting states with opposite helicity. A more recent study~\cite{Yue2024} demonstrated that controlled surface etching of Bi$_2$Se$_3$ (selectively removing Se atoms to form a Bi bilayer) leads to Dirac cone relocation in both real and momentum space and charge transfer between surface layers. If preferential sputtering of Se and/or Te under our conditions similarly creates a Bi-rich surface layer, new surface electronic structure could be introduced. Preferential chalcogen removal during ion bombardment is well-established~\cite{Falcone1981,Sigmund1993}, and while our core-level measurements were inconclusive, a thin ($<1\,\mathrm{QL}$) Bi-rich surface cannot be ruled out. 
In \cite{Aguilery2019} different binary and also one ternary materials of the Bi-based TI-family were investigated with density functional theory calculations including the GW approximation. This study revealed that Bi$_2$Te$_2$Se has a large gap between the valence band maximum and conduction band minimum of about $\SI{0.5}{\electronvolt}$ with a Dirac cone almost perfectly linearly dispersing over more than $\SI{0.6}{\electronvolt}$, but only along the $\overline{\Gamma}-\overline{\text{K}}$ direction. Along the $\overline{\Gamma}-\overline{\text{M}}$ the indirect band gap is around $\SI{0.4}{\electronvolt}$ leading also to a smaller energetic range of a pure linearly dispersing Dirac cone. This potential candidate can be safely excluded, as the scenario of the formation of a Bi$_2$Te$_2$Se-layer by preferential sputtering and the appearance of a linearly dispersing band spanning more than $\SI{0.6}{\electronvolt}$ would also require that all data showing the ALS were taken along the same crystal orientation.

Epitaxial Bi bilayers on topological substrates are known to produce hybridized interface Dirac states~\cite{Klimovskikh2024}, and their Rashba-split states would appear at the $\Gamma$-point with helicity opposite to the substrate TSS. Note that the Rashba states from Bi bilayers always come in pairs for both positive and negative wave vectors~\cite{Yue2024, Klimovskikh2024}, which we could not observe in the case of ALS.

\begin{figure}[t]
\includegraphics[width = 1\columnwidth]{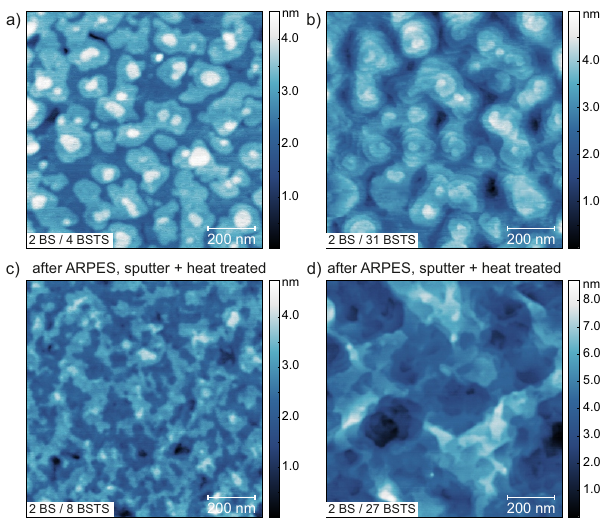}
\caption{a) and b) show typical AFM scans directly after the MBE growth of samples with (2)BS and different thicknesses of the BSTS, a) for $d_\mathrm{BSTS}=4$QL and b) $d_\mathrm{BSTS}=31$QL. After the beamtime and hence after the sputter-heating cycles the roughness and especially the morphology changed, c) and d). c) shows the surface structure of the sample whose ARPES data is presented in Fig.\ref{fig:fig2}~b), and in d) the same for the ARPES data shown in Fig.\ref{fig:fig2}~c) or Fig.\ref{fig:fig1}~b).  }\label{fig:fig3}
\end{figure}

\subsection{High-index surfaces}

High-index Bi(441) surfaces were found to show two Dirac-like states \cite{Bianchi2015}, energetically shifted and with the same helicity as the regular TSS. As the formation of this high-index surface was along one crystal direction, while the perpendicular direction did not show a periodic superlattice, the dispersion relation of the extra band was flat in one direction. This scenario is, however, very unlikely to be the cause for the ALS as the strong intensity and the sharpness would require a high density of periodic potential ripples and a spatially constant potential shift, which is not supported by our AFM scans in Fig.~\ref{fig:fig3}(c,d). The influence of a periodic superstructure on the surface on ARPES data was already discussed in \cite{Himpsel1980}, leading to the generation of extra bands in ARPES. This was also shown \cite{Usachov2012,Polley2019} in the case of Moir\'e pattern introducing complex, but still regular LEED (Low energy electron diffraction) pattern and also to the occurrence of replica bands in ARPES. The latter can also be excluded as replica bands does not extent the energetic range of the original bands. 

Virk and Yazyev~\cite{Virk2016} showed that for high-index surfaces of Bi$_2$Se$_3$ nanostructures the effective bulk band gap confining the Dirac cone can be substantially larger than for the standard (0001) orientation. If the sputter-heating cycle fragments and reorients the outermost quintuple layer to expose local high-index facets, the confining gap for a TSS at these facets could substantially exceed the bulk value, explaining the unusually large energetic range of the ALS. This scenario would lead to electronic and/or topological decoupling of the first few layers. The decoupling could result from an incommensurate lattice of the outermost QL-islands of the material after fragmentation and relaxation due to the reduced lateral extent of the first QL-islands. Hence, the sputter-heating cycle could lead to electronically decoupled nanocrystals with different orientation.

\subsection{Dual topological phase analogy}

Eschbach \textit{et al.}~\cite{Eschbach2017,Morgenstern2021} observed quasi-linearly dispersing, spin-polarized bands spanning $400$--$\SI{600}{\milli\electronvolt}$ in Bi$_1$Te$_1$, which forms as alternating Bi$_2$Te$_3$ QL and Bi bilayers. Density functional theory showed these arise from multiple overlapping spin-polarized bands at non-$\Gamma$ high-symmetry points. However, all our ARPES scans are performed at the $\Gamma$-point and both the ALS and the regular TSS are simultaneously visible, excluding a scenario in which the linearly dispersing feature arises only away from $\Gamma$. The potential formation of a local BiTe-like stoichiometry by the sputter-heating cycle remains an open possibility but cannot be confirmed by our core-level data alone.

\subsection{Material deposition, incorporation and Fermi-velocity}

Although we do not expect material deposition or incorporation during the sputter-heating cycle, literature on this topic~\cite{Wray2011,Nakayama2019,Shvets2017,Virojanadara2010,Xia_2014} sometimes shows supposedly similar results, but with at least one property of extra feature (i) clearly missing: energetic range, linearity, spin-polarization, or material class. Further, the equality of the Fermi velocities of the ALS and the regular TSS $(v_\mathrm{F}^\mathrm{ALS} \approx v_\mathrm{F}^\mathrm{TSS})$ is particularly constraining. Any proposed model must explain this coincidence, which appears non-trivial if the ALS arises from a chemically or structurally distinct surface phase.

\subsection{Outlook}

We emphasize that a definitive assignment of the ALS to a specific mechanism is not possible at present. Identifying its origin will likely require spatially resolved ARPES (nano-ARPES) to determine whether the ALS is spatially uniform or associated with specific surface facets, LEED to characterize the post-treatment surface reconstruction, and first-principles calculations for disordered BSTS surfaces. The application relevance of the ALS, irrespective of its microscopic origin, lies in the substantially enhanced density of states at \EF\ compared to the single regular TSS. 

% ============================================================
\section{Conclusion}
\label{sec:conclusion}

We have observed, following Ar-ion bombardment and annealing of BSTS/Bi$_2$Se$_3$ heterostructures, an anomalous linearly dispersing state (ALS) spanning up to ${\sim}\,\SI{650}{\milli\electronvolt}$ at the $\Gamma$-point. The ALS exhibits spin-momentum locking with a helicity opposite to the regular TSS and a Fermi velocity $(5.1\pm 0.4)\times 10^5\,\frac m s$ indistinguishable from that of the TSS. The effect is reproducible across samples of varying thickness and was confirmed at two independent ARPES facilities. Candidate explanations---sputtering-induced TSS relocation, Bi-bilayer formation via preferential chalcogen removal, and high-index surface reorientation---are discussed but none fully accounts for all observations. We hope that reporting these application-relevant findings will motivate further theoretical and experimental study.

% ============================================================
\begin{acknowledgments}
We acknowledge financial support of the Deutsche  Forschungsgemeinschaft through CRC/SFB 1277 and Project ID 314695032. J.M. thanks the project Quantum Materials for Applications in Sustainable Technologies (QM4ST), funded as Project No.  \verb|CZ.02.01.01/00/22_008/0004572| by Programme Johannes Amos Comenius, through the Excellent Research call.

\end{acknowledgments}

% ============================================================

% for biber:
%\printbibliography

\bibliography{bibJabref}

\end{document}

\title{Supplemental Material: Generation of an anomalous linearly dispersing spin-polarized band in Bi-based topological insulators}

\author{Matthias Kronseder}
\email{matthias.kronseder@ur.de}
\affiliation{Institute for Experimental and Applied Physics,
             University of Regensburg, 93040 Regensburg, Germany}

\author{Thomas Mayer}
\affiliation{Institute for Experimental and Applied Physics,
             University of Regensburg, 93040 Regensburg, Germany}

\author{Jan Min\'{a}r}
\affiliation{New Technologies -- Research Centre of the University of West Bohemia,
             Plze\v{n}, Czech Republic}

\author{Magdalena Marganska}
\affiliation{Institute for Experimental and Applied Physics,
             University of Regensburg, 93040 Regensburg, Germany}

\author{Hedwig Werner}
\affiliation{Institute for Experimental and Applied Physics,
             University of Regensburg, 93040 Regensburg, Germany}

\author{Florian Schmid}
\affiliation{Institute for Experimental and Applied Physics,
             University of Regensburg, 93040 Regensburg, Germany}

\author{Rebeca D\'{i}az-Pardo}
\affiliation{School of Natural Sciences, Department of Physics,
             Technical University of Munich, 85748 Garching, Germany}

\author{Ivana Vobornik}
\affiliation{Elettra-Sincrotrone Trieste, Trieste, Italy}

\author{Jun Fuji}
\affiliation{Elettra-Sincrotrone Trieste, Trieste, Italy}

\author{Cornelia Streeck}
\affiliation{Physikalisch-Technische Bundesanstalt, Berlin, Germany}

\author{Alexander Gottwald}
\affiliation{Physikalisch-Technische Bundesanstalt, Berlin, Germany}

\author{Hendrik Kaser}
\affiliation{Physikalisch-Technische Bundesanstalt, Berlin, Germany}

\author{Bernd K\"{a}stner}
\affiliation{Physikalisch-Technische Bundesanstalt, Berlin, Germany}

\author{Christian H. Back}
\affiliation{School of Natural Sciences, Department of Physics,
             Technical University of Munich, 85748 Garching, Germany}
\affiliation{Center for Quantum Engineering (ZQE),
             Technical University Munich, 85748 Garching, Germany}

\date{\today}

\maketitle

\section{ARPES data from PTB}

\begin{figure}[t]
\includegraphics[width = 1\columnwidth]{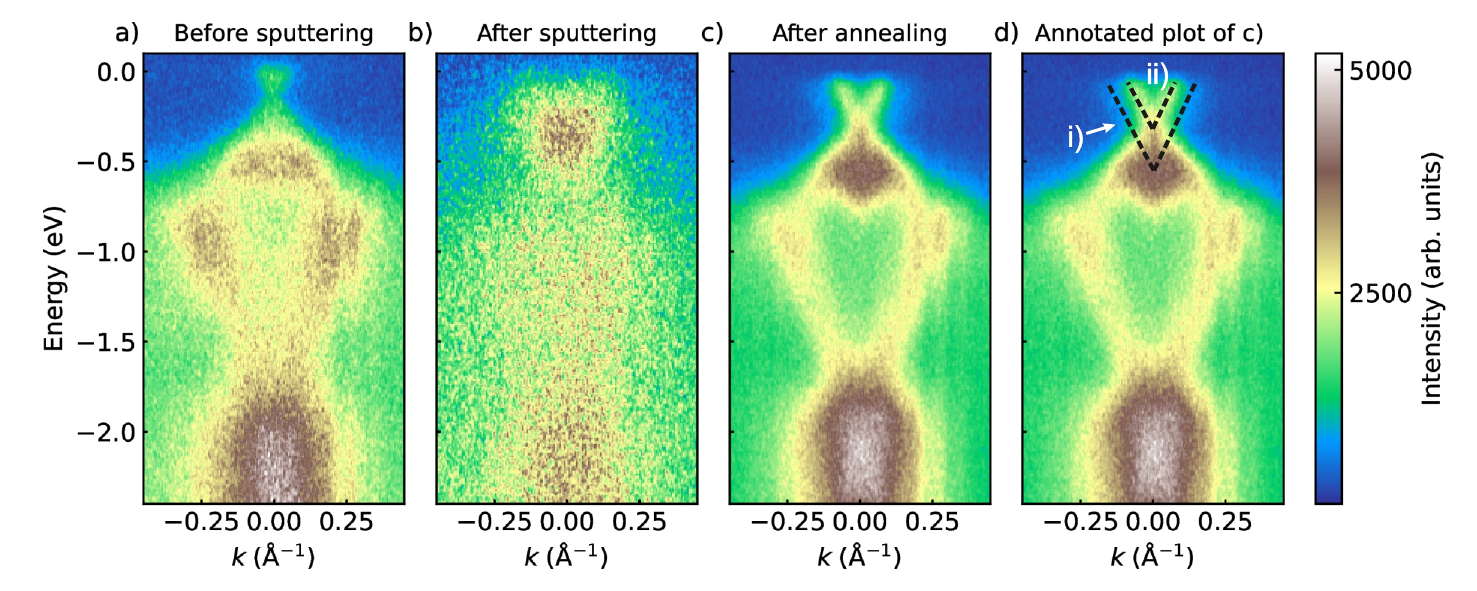}
\caption{ARPES scans performed at the Metrology Light Source of the PTB on a sample with 1BS / 30QL BSTS. a) shows the ARPES scan after thermally decapping the sample. b) after a soft sputtering treatment. c) after another annealing step the ARPES data clearly improved and shows a faint halo around the Dirac cone. d) represents the same data as in c) but with annotations marking the ALS feature with i) and the regular Dirac cone with ii). }\label{fig:figSM1}
\end{figure}

To confirm instrument-independency of the ALS, sample treatment and ARPES measurements were performed at two different synchrotron facilities. Data from the Elettra Synchrotron in Trieste, Italy, are shown in the main manuscript. Data from the Metrology Light Source of the Physikalisch-Technische Bundesanstalt (Berlin, Germany) is shown in Fig.\ref{fig:figSM1} from a sample with the following structure: 
SrTiO$_3$(111) / (1 QL) Bi$_2$Se$_3$ / (30 QL) $({\rm Bi}_{1-x}{\rm Sb}_x)_2({\rm Te}_{1-y}{\rm Se}_y)_3$ / Se-capping. The sputtering-heating cycle was performed in a similar way with respect to the treatment at the Elettra Synchrotron. The sample used at the PTB has a slightly thicker BSTS and the capping was pure Selenium instead of Tellurium/Selenium. After decapping by thermal treatment and a first bombardment with Ar-ions at 400V for 10 minutes, the ARPES scans get blurred, Fig. S1 b). After annealing at 190C for 20 minutes, the band structure shows up again with larger contrast than before the treatment. A faint structure not seen before appears around \EF\ and close to the Dirac cone. Although the contrast of this feature is very low and assuming a linear dispersion, the point where the two lines merge is around -0.5eV.

\bibliography{bibJabref}